# How do you define and measure research productivity?[1]

*Giovanni Abramo*[a,*], *Ciriaco Andrea D'Angelo*[b,a]

[a] Laboratory for Studies of Research and Technology Transfer
Institute for System Analysis and Computer Science (IASI-CNR)
National Research Council of Italy

[b] Department of Engineering and Management
University of Rome "Tor Vergata" - Italy

**Abstract**

Productivity is the quintessential indicator of efficiency in any production system. It seems it has become a norm in bibliometrics to define research productivity as the number of publications per researcher, distinguishing it from impact. In this work we operationalize the economic concept of productivity for the specific context of research activity and show the limits of the commonly accepted definition. We propose then a measurable form of research productivity through the indicator "Fractional Scientific Strength (FSS)", in keeping with the microeconomic theory of production. We present the methodology for measure of FSS at various levels of analysis: individual, field, discipline, department, institution, region and nation. Finally, we compare the ranking lists of Italian universities by the two definitions of research productivity.

**Keywords**

*Research productivity; FSS; research evaluation; university rankings.*

**Introduction**

In 2010, Opthof and Leydesdorff criticized the statistical normalization of the Leiden CWTS "crown indicator". A year later, bibliometricians from the CWTS group (Waltman et al. 2011) admitted that the "old crown indicator" was mathematically inconsistent and adopted the normalization method suggested by the above authors, leading to a "new crown indicator": the mean normalized citation score, or MNCS. A counter-reply from Leydesdorff and Opthof (2011) was not long in arriving: although agreeing with the new statistical normalization, they then further recommended using the mean rather than the median to field normalize citations.

In a parallel story, since the original introduction of the h-index in 2005 by physicist Jorge E. Hirsch, over 1,600 articles have been written illustrating its merits and defects and proposing one variant after another, to the extent even the most devoted historian of bibliometrics would despair of tracing them all.

But is it possible that these two research performance indicators really merited all this attention, or is it a case of "Much ado about nothing"? These particular indicators have only been the most popular among a myriad of others proposed over recent years by scholars and practitioners. While bibliometricians undoubtedly intended to provide useful indicators and ever more accurate and reliable methods, they have actually been the cause of increasing confusion. The proliferation of proposals has actually generated a type of disorientation among decision makers, no longer able to discriminate the pros and cons of the various indicators for planning an actual evaluation exercise. The proof of this is the increasing number of expert commissions and working groups at institutional, national and supranational levels, formed to deliberate and recommend on this indicator, that set of indicators, and this or that measure of performance. Performance ranking lists at national and international levels are published with media fanfare, influencing opinion and practical choices. The impression of the current authors is that these rankings of scientific performance, produced by "non-bibliometricians" (THE 2013; SJTU 2013; QS 2013; etc.) and even by bibliometricians (University of Leiden, SCImago, etc.), are largely based on what can easily be counted rather than "what really counts". It is also our impression that the large part of the performance evaluation indicators proposed in the literature arise from a primarily mathematical school of thought. While knowledge in this area is fundamental in the methodology for application, our personal conviction is that research evaluation indicators must necessarily derive from economic theory. Since research activity is a production process, it should be analyzed from the perspective of microeconomic theory of production. Performance, or the ability to perform, should be evaluated with respect to the specific goals and objectives to be achieved. Because objectives may vary across research institutions and along time, recommending a sole indicator of performance would be inappropriate. However this does not justify the proliferation of hundreds of indicators, which while they may offer ease of calculation have little or no utility for policy or management decisions. A nation or research institution could aim to improve its average research productivity, or the rate of concentration of top articles or top scientists, or aim for a combination of these efforts in different weights. However we doubt that any government or research administrator could pursue any improvement that would be revealed through measurement by the MNCS, h-index or average number of publications per researcher. As a consequence, all the research evaluations based on these indicators and their relative rankings are at best of little or no value, and are otherwise actually dangerous, due to the distortions embedded in the information provided to the decision-makers. What is certain is that the objectives for research systems must be stated in measurable terms representing the desired outcome of production activity.

In this work we intend to provide an operative definition of the principal indicator of efficiency of any production unit, productivity. In the context of research organizations, bibliometricians have become accustomed to define research productivity as the number of publications per researcher, distinguishing it from impact, which they measure by citations. Honestly, we are not able to date back the scholar who first introduced the above definition, but already in 1926 Alfred J. Lotka used the number of publications in his milestone work (Lotka 1926) where he presented what it is now known as the Lotka's law or research productivity. Unfortunately, from an economic standpoint, and we remind that productivity is a concept born into the economic theory, such definition makes little sense. It would be acceptable only if all publications had the same value or impact, but that could not be further from the truth.

The objective of this paper is to operationalize the concept of productivity for the specific context of research activity and propose a measurable form of productivity. We will then present an indicator, Fractional Scientific Strength (FSS), which in our view is thus far the best in approximating the measure of productivity. We will also illustrate the methodology for measuring FSS in the evaluation of performance at various levels of analysis: individual, field, discipline, department, institution, region and nation. Finally, we compare Italian university ranking lists by the two definitions of productivity: FSS and average number of publications per researcher.

**Productivity in research activities**

In this section, our intention is to operationalize the concept of research productivity in simple terms and propose a proxy to measure it.

Generally speaking, the objective of research activity is to produce new knowledge. Research activity is a production process in which the inputs consist of human, tangible (scientific instruments, materials, etc.) and intangible (accumulated knowledge, social networks, economic rents, etc.) resources, and where output, the new knowledge, has a complex character of both tangible nature (publications, patents, conference presentations, databases, etc.) and intangible nature (tacit knowledge, consulting activity, etc.). The new-knowledge production function has therefore a multi-input and multi-output character. The principal efficiency indicator of any production unit (individual, research group, department, institution, field, country) is productivity: in simple terms the output produced in a given period per unit of production factors used to produce it. To calculate research productivity one needs adopt a few simplifications and assumptions.

On the output side, a first approximation arrives from the imposition of not being able to measure any new knowledge that is not codified. Second, where new knowledge is indeed codified, we are faced with the problem of identifying and measuring its various forms. It has been shown (Moed 2005) that in the so-called hard sciences, the prevalent form of codification for research output is publication in scientific journals. Such databases as Scopus and Web of Science (WoS) have been extensively used and tested in bibliometric analyses, and are sufficiently transparent in terms of their content and coverage. As a proxy of total output in the hard sciences, we can thus simply consider publications indexed in either WoS or Scopus[2]. With this proxy, those publications that are not censused will inevitably be ignored. This approximation is considered acceptable in the hard sciences, although not for the arts, humanities and a good part of the social science fields. Other forms of output, particularly patents, can be identified in commercial or free databases such as Derwent and

---

[2] Although the overall coverage of the two databases does differ significantly, evidence suggests that, with respect to comparisons at large scale level in the hard sciences, the use of either source yields similar results (Archambault et al. 2009).

Espacenet. Patents are often followed by publications that describe their content in the scientific arena, so the analysis of publications alone may actually avoid in many cases a potential double counting.

Research projects frequently involve a team of researchers, which shows in co-authorship of publications. Productivity measures then need to account for the fractional contributions of single units to outputs. The contributions of the individual co-authors to the achievement of the publication are not necessarily equal, and in some fields the authors signal the different contributions through their order in the byline. The conventions on the ordering of authors for scientific papers differ across fields (Pontille 2004; RIN 2009), thus the fractional contribution of the individuals must be weighted accordingly. Following these lines of logic, all performance indicators based on full counting or "straight" counting (where only the first author or the corresponding author receive full credit and all others receive none) are invalid measures of productivity. The same invalidity applies to all indicators based on equal fractional counting in fields where co-author order has recognized meaning.

Furthermore, because the intensity of publications varies across fields (Garfield 1979; Moed et al. 1985; Butler 2007), in order to avoid distortions in productivity rankings (Abramo et al. 2008), one should compare researchers within the same field. A prerequisite of any productivity assessment free of distortions is then a classification of each individual researcher in one and only one field. An immediate corollary is that the productivity of units that are heterogeneous for fields of research of their staff cannot be directly measured at the aggregate level, and that there must be a two-step procedure: first measuring the productivity of the individual researchers in their field, and then appropriately aggregating this data.

In bibliometrics we have seen the evolution of language where the term "productivity" measures refers to those based on publication counts while "impact" measures are those based on citation counts. In a microeconomic perspective, the first operational definition would actually make sense only if we then compare units that produce output of the same value. In reality this does not occur, because the publications embedding the new knowledge produced have different values. Their value is measured by their impact on scientific advancements. As proxy of impact bibliometricians adopt the number of citations for the units' publications, in spite of the limits of this indicator (negative citations, network citations, etc.) (Glänzel 2008). Citations do in fact demonstrate the dissemination of knowledge, creating conditions for knowledge spillover benefits. Citations thus represent a proxy measure of the value of output. To each citation may also be given a different weight, depending on the citing article influence, measured by number of citations accrued.

Comparing units' productivity by field is not enough to avoid distortions in rankings. In fact citation behavior too varies across fields, and is not unlikely that researchers belonging to a particular scientific field may also publish outside that field (a typical example is statisticians, who may apply statistics to medicine, physics, social sciences, etc.). For this reason bibliometricians standardize the citations of each publication with respect to a scaling factor stemming from the distribution of citations for all publications of the same year and the same subject category.[3] Different scaling factors have been suggested and adopted to field normalize citations (average, median, z-score of normalized distributions, etc.). Because interdisciplinary work may easily suffer in the evaluation from being misplaced in a categorical classification system (Laudel & Origgi 2006), few scholars have proposed to normalize citations by the number of bibliographic references of the citing paper (Pepe & Kurtz 2012; Leydesdorff & Bornmann 2011).

On the side of production factors, there are again difficulties in measure that lead to inevitable approximations. The identification of production factors other than labor and the calculation of their value and share by fields is formidable (consider quantifying value of accumulated knowledge or scientific instruments shared among units). Furthermore, depending on the objectives of the assessment exercise, it could sometimes be useful to isolate and examine the contribution to output of factors, that are independent of the capacities of the staff for the units under examination (for example returns to scale, returns to scope, available capital, etc.).

---

[3] The subject category of a publication corresponds to that of the journal where it is published. For publications in multidisciplinary journals the scaling factor is generally calculated as the average of the standardized values for each subject category.

**Total factor research productivity**

The productivity of the total production factors is therefore not easily measurable. There are two traditional approaches used by scholars to measure the total factor productivity: parametric and non-parametric techniques.

Parametric methodologies are based on the a priori definition of the function that can most effectively represent the relationship between input and output of a particular production unit. These estimation processes have the purpose of determining the coefficients (model parameters) of a regression equation describing the production function, usually a Cobb-Douglas type equation. The main limitation of such methodology concerns the need for advance definition of closed models describing the production function: this entails the need to make assumptions on the relationship between input and output, for instance to assume additive inputs rather than a linear function connecting the two values. Furthermore, parametric techniques cannot identify benchmark best practices, but define expected (or optimal) performances at selected input levels.

The purpose of non-parametric methods, on the other hand, is to compare empirically measured performances of production units (commonly known as Decision Making Units, DMUs), in order to define an "efficient" production frontier, comprising the most productive DMUs. The reconstruction of that frontier is useful to assess the inefficiency of the other DMUs, based on minimum distance from the frontier. The main advantages of non-parametric methods can be summarized as follows:
- Complex production systems with multiple inputs and outputs are assessed by means of a single global efficiency value, the Total Factor Productivity, obtained with no pre-defined weighting factors of any sort;
- No functional relationship needs to be established to define production processes, nor do optimization or estimation processes;
- The frontier from which efficiency coefficients are calculated is obtained from actually measured DMUs - in other words, comparisons are to real production units that are used as references for best practices.

For both methodologies, correct identification of inputs and output indicators is crucial to the reliability of the model application.

Given the characteristics of the two models described, the non-parametric approach is generally preferable in the context of new knowledge production.

One of the non-parametric methods most commonly observed in the literature is the DEA. The DEA was developed as a technique for assessing the efficiency of industrial production systems (Charnes et al. 1978; Banker et al. 1984) and has extremely limited applicability hypotheses: i) homogeneity of DMUs - the production units must produce the same type of goods or services using the same type of resources; ii) convexity of the analyzed set - the frontier includes all possible linear combinations of the efficient units; iii) free disposability, meaning the possibility to eliminate resources with no costs.

There are two DEA application models: output-oriented and input-oriented. In the former,

the efficiency deviation from the frontier is evaluated as the maximum equiproportional increase of all outputs allowed by the available inputs. This model is particularly appropriate for scientific research, since in general the overall objective is not to reduce the input while maintaining constant production, but to maximize production with the resources available. The DEA methodology includes two distinct models for cases of absence (CRS) or presence of returns to scale of production factors (VRS). The use of the CRS specification when not all DMUs are operating at optimal scale, will result in measures of *technical efficiency* (TE) which are confounded by *scale efficiency* (SE). The use of the VRS specification will permit the calculation of TE devoid of these SE effects. The SE can be extracted by applying both models to the same data set. The problem of calculating the frontier and the DEA efficiency indexes can be formulated in terms of linear programming and is easily solved by using specially developed software, such as the Efficiency Measurement System (EMS) developed by the University of Dortmund (Scheel 2000). The use of the DEA method should, in any case, be supported by technical-methodological comments which can help correctly interpret any results arising out of it. First, the DEA is of purely deterministic nature: any deviation from the frontier is associated with inefficiency, and it is not possible to take into consideration casual elements or external noise which might have affected the results. Secondly, the calculated efficiency measure is only valid for the variables that are measured and used by the model. While representing measures of total productivity, those values depend exclusively from the choice of variables, and might therefore not give a completely representative picture of the efficiency of DMUs, especially as important input or output factors could be overlooked. In the specific case of the bibliometric-type measurement of the production performance of Universities with the DEA model, possible distortions might, for instance, arise if: (on the input side) time is allocated incongruously between research and teaching or between different types of research (basic/applied), or production factors overlooked in the model are non-homogenously available, such as scientific instruments, or non-employed staff (PhD students, external collaborators); (on the output side) researchers have different inclinations to codify their results under forms other than publication, or there are divergent agglomeration[4] or scope economies.

The measure of total factor productivity requires information on the different production factors by unit of analysis. Instead of total factor research productivity, most often research administrators are interested in measuring and comparing simply labor productivity, i.e. output per unit value of labor, all other production factors being equal. The next section describes the requirements for this kind of analysis.

**Labor productivity in research activity and the FSS**

In measuring labor productivity then, if there are differences of production factors available to each unit, one should normalize for these. Unfortunately, relevant data are not easily available, especially at the individual level. Thus an often-necessary assumption is that the resources available to units within the same field are the same. A further assumption, again unless specific data are available, is that the hours devoted to research are more or less the same for each individual. Finally, as occurs for output, the value of researchers is not undifferentiated and this is reflected in the different cost of labor, which varies among research staff, both within and between units. To measure the cost efficiency or research units, one should normalize its output by the cost of labor. In a study of Italian universities, Abramo et al. (2011) demonstrated that productivity of full, associate and assistant professors is different. Because academic rank in general determines differentiation in salaries, if information on individual salaries in unavailable, then one can still reduce the distortion in productivity measures by differentiating performance rankings by academic rank.

---

[4] A host of studies have demonstrated the positive effect of proximity of private research on the research productivity of public laboratories (Siegel et al. 2003).

Next we propose our best proxy for the measurement of the average yearly labor productivity at various unit levels (individual, field, discipline, department, entire organization, region and country). The indicator is FSS, which we have previously applied to the Italian higher education context, where most of its embedded approximations and assumptions are legitimate.

As noted above, for any productivity ranking concerning units that are non-homogenous for their research fields, it is necessary to start from the measure of productivity of the individual researchers or fields. Without these two building blocks, any measure at aggregate level presents strong distortions (Abramo et al. 2008). In their measures of this data, the authors gain advantage from a characteristic that seems unique to the Italian higher education system, in which each professor is classified as belonging to a single research field. These formally-defined fields are called "Scientific Disciplinary Sectors" (SDSs): there are 370 SDSs, grouped into 14 "University Disciplinary Areas" (UDAs). In the hard sciences, there are 205 such fields[5] grouped into nine UDAs.[6]

When measuring research productivity, the specifications for the exercise must also include the publication period and the "citation window" to be observed. The choice of the publication period has to address often contrasting needs: ensuring the reliability of the results issuing from the evaluation, but also permitting conduct of frequent assessments. For the most appropriate publication period to be observed see Abramo et al. (2012a), while for the citation window that optimizes the tradeoff between accuracy of rankings and timeliness of the evaluation exercise, see Abramo et al. (2012b).

*Labor productivity at the individual level*

At micro-unit level (the individual researcher level, $R$) we measure $FSS_R$, a proxy of the average yearly productivity over a period of time, accounting for the cost of labor. In formula:

$$FSS_R = \frac{1}{w_R} \cdot \frac{1}{t} \sum_{i=1}^{N} \frac{c_i}{\bar{c}} f_i$$

[1]

Where:
$w_R$ = average yearly salary of the researcher[7];
t = number of years of work of the researcher in the period of observation;
N = number of publications of the researcher in the period of observation;
$c_i$ = citations received by publication $i$;
$\bar{c}$ = average of the distribution of citations received for all cited publications[8] of the same year and subject category of publication $i$;
$f_i$ = fractional contribution of the researcher to publication $i$.

---

[5] The complete list is accessible on http://attiministeriali.miur.it/UserFiles/115.htm, last accessed on Feb. 13, 2014.
[6] Mathematics and computer sciences; physics; chemistry; earth sciences; biology; medicine; agricultural and veterinary sciences; civil engineering; industrial and information engineering.
[7] We assume that other production factors are equally available to all researchers. If not, their value should be taken into account.
[8] A preceding article by the same authors demonstrated that the average of the distribution of citations received for all cited publications of the same year and subject category is the most effective scaling factor (Abramo et al. 2012c).

Fractional contribution equals the inverse of the number of authors, in those fields where the practice is to place the authors in simple alphabetical order, but assumes different weights in other cases. For the life sciences, widespread practice in Italy and abroad is for the authors to indicate the various contributions to the published research by the order of the names in the byline. For these areas, we give different weights to each co-author according to their order in the byline and the character of the co-authorship (intra-mural or extra-mural). If first and last authors belong to the same university, 40% of citations are attributed to each of them; the remaining 20% are divided among all other authors. If the first two and last two authors belong to different universities, 30% of citations are attributed to first and last authors; 15% of citations are attributed to second and last author but one; the remaining 10% are divided among all others[9]. Failure to account for the number and position of authors in the byline would result in notable ranking distortions both at the individual (Abramo et al. 2013a), and aggregate (Abramo et al. 2013b) levels.

To calculate productivity accounting for the cost of labor, requires knowledge of the cost of each researcher, information that is usually unavailable for reasons of privacy. In the Italian case we have resorted to a proxy. In the Italian university system, salaries are established at the national level and fixed by academic rank and seniority. Thus all professors of the same academic rank and seniority receive the same salary, regardless of the university that employs them. The information on individual salaries is unavailable but the salaries ranges for rank and seniority are published. Thus we have approximated the salary for each individual as the national average of their academic rank.

If information on salary is not available at all, one should at least compare research performance of individuals of the same academic rank. Failure to account for the cost of labor would result in ranking distortions as shown by Abramo et al. (2010).

We calculate the productivity of each scientist in each SDS and express it on a percentile scale of 0-100 (worst to best) for comparison with the performance of all Italian colleagues of the same SDS; or as the ratio to the average productivity of all Italian colleagues of the same SDS with productivity above zero[10]. In general we can exclude, for the Italian case, that productivity ranking lists may be distorted by variable returns to scale, due to different sizes of universities (Abramo et al. 2012d) or by returns to scope of research fields (Abramo et al. 2013d).

*Labor productivity in a specific field*

At field level $S$, the yearly average productivity $FSS_S$ over a certain period for researchers in a university (region, country, etc.) in a specific SDS[11] is:

---

[9] The weighting values were assigned following advice from senior Italian professors in the life sciences. The values could be changed to suit different practices in other national contexts.

[10] In a preceding article the authors demonstrated that the average of the productivity distribution of researchers with productivity above 0 is the most effective scaling factor to compare the performance of researchers of different fields (Abramo et al. 2013c).

[11] We note again that a field is not an organizational unit, rather a classification of researchers by their scientific qualifications. This does not mean that all the researchers in the same field and organization will necessarily form a single research group that works together. As an example, we quote the SDS description for FIS/03- Condensed matter physics: "The sector includes the competencies necessary for dealing with theory and experimentation in the state of atomic and molecular aggregates, as well as competencies suited to dealing with properties of propagation and interaction of photons in fields and with material. Competencies in this sector also concern research in fields of atomic and molecular physics, liquid and solid states, semiconductors and metallic element composites, dilute and plasma states, as well as photonics, optics, optical electronics and quantum electronics". In the Italian academic system it is quite common to find "Condensed matter physics" researchers working in two different departments (physics and engineering) at the same university.

$$FSS_S = \frac{1}{w_S} \sum_{i=1}^{N} \frac{c_i}{\bar{c}} f_i$$

[2]

Where:
$w_S$ = total salary of the research staff of the university in the SDS, in the observed period;
$N$ = number of publications of the research staff in the SDS of the university, in the period of observation;
$c_i$ = citations received by publication $i$;
$\bar{c}$ = average citations received by all cited publications of the same year and subject category of publication $i$;
$f_i$ = fractional contribution of researchers in the SDS of the university, to publication $i$, calculated as described above.

For each SDS we can construct a university (region, country, etc.) productivity ranking list by $FSS_S$ expressed in percentiles or as the $FSS_S$ ratio to average $FSS_S$ of all universities with productivity above zero in the SDS.

The measures of productivity at field level permit identification of field strengths and weaknesses and thus correctly inform research policies and strategies.

*Labor productivity of multi-fields units*

In multi-field organizational units (i.e. disciplines, departments, universities, regions, nations), where there are researchers that belong to different fields, we are presented with the problem of how to aggregate productivity measures for researchers from the various fields. Two methods are possible, based on either the performance of individual researchers ($FSS_R$), or of the SDSs ($FSS_S$) present in the unit under examination. The appropriate choice depends on the objective for the measure. The first method emphasizes individual performance while the second emphasizes field performance, which we note is a "virtual" unit, since the members of the SDS at a university do not necessarily work together on a structured basis. The research administrator will perhaps be more interested in the performance results derived under the first method, determined from the average of individual productivities. On the other hand the policy-maker, not being particularly interested in the performance variability within the organizational units but rather in comparison of the overall productivity of the various research institutions, could prefer the performance measure calculated by the second method. In the following subsections we present the two measurement procedures.

Labor productivity of multi-fields units based on FSS$_R$

We have seen that the performance of the individual researchers in a unit can be expressed in percentile rank or standardized to the field average. The natural tendency would be to express the productivity of multi-field units by the simple average of the percentile ranks of the researchers. It should be noted though that the resort to percentile rank for the performance measure in multi-filed units or for simple comparison of performance for researchers in different fields is subject to obvious limitations, the first being compression of the performance differences between one position and the next. Thompson (1993) warns that percentile ranks should not be added or averaged, because percentile is a numeral that does not represent equal-interval measurement. Further, percentile rank is also sensitive to the size of the fields and to the performance distribution. For example, consider a unit composed of two researchers in two different SDSs (A and B, each with a national total of 10 researchers), who both rank in third place, but both with productivity only slightly below that of the first-ranked researchers in their respective SDSs: the average rank percentile for the unit will be 70. Then consider another unit with two researchers belonging to another two SDSs (C and D, each with 100 researchers), where both of the individuals place third but now with a greater gap to the top scientists of their SDSs: their percentile rank will be 97. In this particular example, a comparison of the two units using percentile rank would certainly penalize the former unit.

The second approach, instead involving standardization of productivity by field average, takes account of the extent of difference between productivities of the individuals. In formula, the productivity $FSS_D$ over a certain period for department $D$, composed of researchers that belong to different SDSs:

$$FSS_D = \frac{1}{RS} \sum_{j=1}^{RS} \frac{FSS_{R_j}}{\overline{FSS_R}}$$

[3]

Where:
$RS$ = research staff of the department, in the observed period;
$FSS_{R_j}$ = productivity of researcher $j$ in the department;
$\overline{FSS_R}$ = national average productivity of all productive researchers in the same SDS of researcher $j$.

Labor productivity of multi-fields units based on FSS$_S$

The second method for measurement of research unit productivity involves identifying all the SDSs present in the unit and assigning each one a relative weight depending on size (full time equivalent research personnel). As an example, for measurement of productivity of a university (region, nation) in a discipline (UDA), beginning from the productivity of the individual SDSs (FSSs), the productivity $FSS_U$ of a university in a specific UDA $U$, is:

$$FSS_U = \sum_{k=1}^{N_U} \frac{FSS_{S_k}}{\overline{FSS_{S_k}}} \frac{w_{S_k}}{w_U}$$

[4]

With:
$w_{S_k}$ = total salary of the research staff of the university in the SDS $k$, in the observed period;
$w_U$ = total salary of the research staff of the university in the UDA $U$, in the observed period;
$N_U$ = number of SDSs of the university in the UDA $U$;
$\overline{FSS_{S_k}}$ = weighted[12] average $FSS_S$ of all universities with productivity above 0 in the SDS $k$.

For the measure of the productivity of a department (or university, region, country), the procedure is exactly the same: the only thing that changes is the size weight of the SDS, which is no longer with respect to the other SDSs of the UDA, but rather to all the SDSs of the department (university, region, country).

As noted, the appropriate choice between the two methods of measure for performance of a multi-field unit depends on the aims of the evaluation. The first method, based on productivity of individual researchers, interprets the performance of the unit as the average of the individual performances, meaning that the emphasis is on the individual. The other method, based on productivity of fields, interprets the field as a unique group (even though a virtual group), meaning that emphasis is on the overall product of the researchers that belong to the field, independently of the variability of the individual contributions. The two methods lead to performance results that are quite similar. In a future work we will provide a comparative in-depth analysis of the two methods.

---

[12] The weight represents the relative size (in terms of cost of labor) of the SDS of each university.

*Comparison of university ranking lists based on different research productivity measures*

In this section we compare the Italian university ranking lists by *FSS* with those by the commonly accepted definition of productivity, i.e. number of publications per researcher, which we call *P*. We also compare ranking lists by *FSS* with those by the main variant of *P*, embedding fractional counting for co-authored publications, which we call *FP*. The production period under observation is 2006-2010, while citations are counted on 31/12/2011.

Similar to [1], the number of publications achieved in the observed period is standardized for years of work over the observation period. The individual measurements are then aggregated at the UDA level, through the same standardizations illustrated in [3]:

$$P_U = \frac{1}{RS_U} \sum_{j=1}^{RS_U} \frac{Q_j}{\overline{Q}}$$

[5]

Where:
$RS_U$ = research staff of the university in the UDA *U*, in the observed period;
$Q_j$ = average annual output of researcher *j*, in the observed period;
$\overline{Q}$ = average annual output of all productive national researchers in the same SDS of researcher *j*, in the observed period.

We can now construct university ranking lists per UDA by FSS and P. Table 1 presents descriptive statistics about comparison of rankings. To make our measures more robust we have excluded researchers with less than three years of work in the observed period and universities with research staff below ten units in the UDA.

*Table 1: Comparison of university ranking lists by P and FSS, per UDA*

| UDA | No. of universities | % shifting rank | Average shift | Median shift | Max shift | Correl. | From top to non top quartile |
|---|---|---|---|---|---|---|---|
| Mathematics and computer science | 50 | 98.0% | 7.1 | 5.5 | 28 | 0.781 | 38.5% |
| Physics | 43 | 93.0% | 10.7 | 10 | 34 | 0.426 | 81.8% |
| Chemistry | 41 | 97.6% | 6.7 | 5 | 25 | 0.726 | 45.5% |
| Earth sciences | 30 | 80.0% | 3.7 | 3 | 16 | 0.822 | 25.0% |
| Biology | 49 | 93.9% | 6.2 | 5 | 20 | 0.833 | 23.1% |
| Medicine | 42 | 90.5% | 4.1 | 3.5 | 16 | 0.903 | 36.4% |
| Agricultural and veterinary sciences | 27 | 85.2% | 2.8 | 2 | 10 | 0.868 | 28.6% |
| Civil engineering | 35 | 88.6% | 3.8 | 3 | 11 | 0.878 | 33.3% |
| Industrial and information engineering | 42 | 90.5% | 7.0 | 5 | 28 | 0.682 | 27.3% |
| Total | 61 | 85.2% | 4.9 | 4 | 14 | 0.933 | 18.8% |

The last row shows values referring to all Italian universities without distinction per UDA[13]. The correlation between these rankings is clearly very high: the Spearman coefficient of correlation is equal to 0.933; however a full 52 of the 61 universities evaluated (85%) change position between the two rankings, with an average shift equal of 4.9 positions and the median at 4. One university jumps 14 positions, moving from twentieth place in the FSS ranking to thirty-fourth place in the ranking by P. Three out of 16 universities that placed in the first quartile for FSS finish in the second quartile for P.

---
[13] In this case, universities with research staff in the hard sciences below 30 units were not considered.

The analysis by UDA offers interesting insights. The Medicine UDA shows the highest level of correlation between the two ranking lists (Spearman coefficient 0.903). However the Agricultural and veterinary sciences discipline is the one with the smallest shifts in position: although only 27 universities are evaluated, 23 shows shifts in rank between the two rankings; however the average shift is just 2.8 positions, with median 2 and maximum shift of 10. At the opposite extreme is Physics: here the correlation between the two rankings is low (Spearman coefficient 0.426), with an average shift of 10.7 positions and median of 10. There is even a case of a university that jumps 34 positions, moving from forty-first for FSS to seventh for P. Among the 11 universities at the top for FSS, only two remain "top" for P. In reality, this UDA presents an unquestionable anomaly: particularly in the fields of "Particle and high-energy physics", research is often conducted through so-called "grand experiments". The results typically have high scientific impact and are accredited to a large part of the research staff of the partner organizations. They are disseminated through publications with hundreds or even thousands of co-authors. Thus the fractionalization of the author contribution in FSS gives productivity scores quite different from those arising from P and resulting diverging ranking lists.

We now proceed to a further analysis that considers fractional counting, as defined in [1] in place of full counting of coauthored publications. In formulae:

$$FP_U = \frac{1}{RS_U} \sum_{j=1}^{RS_U} \frac{FQ_j}{\overline{FQ}}$$

[6]

Where:
$RS_U$ = research staff of the university in the UDA $U$, in the observed period;
$FQ_j$ = average annual fractional output of researcher $j$, in the observed period;
$\overline{FQ}$ = average annual fractional output of all productive national researchers in the same SDS of researcher $j$, in the observed period.

As expected, the correlation between FP and FSS ranking lists is stronger than between P and FSS; the number of shifts in each UDA is lower, as well as the number of universities dropping from top quartile.

*Table 2: Comparison of university ranking lists by FP and FSS, per UDA*

| UDA | No. of universities | % shifting rank | Average shift | Median shift | Max shift | Correl. | From top to non top quartile |
|---|---|---|---|---|---|---|---|
| Mathematics and computer science | 50 | 96.0% | 7.1 | 5 | 24 | 0.789 | 23.1% |
| Physics | 43 | 97.7% | 6.1 | 4 | 29 | 0.761 | 18.2% |
| Chemistry | 41 | 87.8% | 6.4 | 6 | 21 | 0.742 | 36.4% |
| Earth sciences | 30 | 86.7% | 3.9 | 3 | 12 | 0.822 | 25.0% |
| Biology | 49 | 85.7% | 4.6 | 3 | 19 | 0.897 | 23.1% |
| Medicine | 42 | 88.1% | 3.4 | 3 | 16 | 0.929 | 27.3% |
| Agricultural and veterinary sciences | 27 | 88.9% | 3.7 | 2 | 13 | 0.784 | 28.6% |
| Civil engineering | 35 | 82.9% | 3.8 | 3 | 14 | 0.873 | 22.2% |
| Industrial and information engineering | 42 | 95.2% | 6.5 | 5 | 27 | 0.654 | 27.3% |
| Total | 61 | 82.0% | 4.0 | 3 | 14 | 0.952 | 18.8% |

**Discussion and conclusions**

Until now, bibliometrics has proposed indicators and methods for measuring research performance that are largely inappropriate from a microeconomics perspective. The h-index and most of its variants, for example, inevitably ignore the impact of works with a number of citations below h and all citations above h of the h-core works. The h-index also fails to field-normalize citations, to account for the number of co-authors and their order in the byline, Last but not least, because of the different intensity of publications across fields, productivity rankings need to be carried out by field (Abramo and D'Angelo 2007), when in reality there is a human tendency to compare h-indexes for researchers across different fields. Each one of the proposed h-variant indicators tackles one of the many drawbacks of the h-index while leaving the others unsolved, so none can be considered completely satisfactory.

The new crown indicator, on the other hand, measures the average standardized citations of a set of publications, which cannot provide any indication of unit productivity. In fact a research unit with double the MNCS value of another unit could actually have half the productivity, if the second unit produced four times as many publications. Whatever the CWTS research group (Waltman et al. 2012) might claim for them, the annual world university rankings by MNCS are not "performance" rankings - unless someone abnormally views performance as average impact of product, rather than impact per unit of cost. Applying the CWTS method, a unit that produces only one article with 10 citations has better performance than a unit producing 100, where each but one of these gets 10 citations and the last one gets nine citations. From a different standpoint, an organization may worsen its MNCS ranking if it produces an additional article whose normalized impact is below the previous MNCS value, which is a paradox. Further, the methodology reported for producing the ranking lists does not describe any weighting for co-authorship on the basis of byline order. Similar drawbacks are embedded is the SCImago Institutions Ranking by their main indicator, the Normalized Impact, measuring the ratio between the average scientific impact of an institution and the world average impact of publications of the same time frame, document type and subject area. We do not further consider any of the many annual world institutional rankings that are severely size dependent: the SJTU Shanghai Jiao Tong University, THE-Times Higher Education and QS Quacquarelli Symonds rankings, among others. These seem to represent skilled communications and marketing operations, with the actual rankings resulting more from improvisation than scientifically-reasoned indicators and methods. Gross attempts to compare the research productivity of nations can be found in the groundbreaking work by May (1997). He measured the relative international standing of 15 countries in science, medicine and engineering, by their shares of ISI-indexed publications and citations. The USA invariably ranked at the top for such indicators in all scientific sectors In an attempt to separate the effect of size (labor and capital) from the effect of the quality of the production factors, he then ranked countries by citations per unit of spending. King (2004) updated May's original work to 2002, covering a 10-year period. The new study increased the number of nations analyzed (31), provided a longitudinal analysis over two five-year periods, added further indicators (top 1% highly cited articles; average citations per paper), provided for normalization of citations to the mean for each field, and took account of year of publication, thus providing aggregate measures of the overall research standing of each country. Outputs and outcomes were then normalized to inputs (researchers, expenditures, GDP). Such attempts to normalize outputs and outcomes to inputs, have dealt with the data at the aggregate level and have not been able to avoid the consequent distortions. In fact while most scholars now typically normalize the observed output and citation data, accounting for the field and year of publication, the data on input are not correspondingly divided according to the fields of allocation, since the practitioners lack data on the numbers of researchers and the expenditures per field in the individual countries under comparison.

Pepe and Kurtz (2012) have proposed a productivity indicator for individuals, the research impact quotient, which has similarities with the FSS. It is the quotient of the square root of "total research impact" to the time of production. The total research impact normalizes external (non-self) citations by the number of co-authors of the cited paper and the number of bibliographic references of the citing paper. While the elimination of self-citations is an acceptable option, the normalization by the number of references of the citing papers is questionable and would deserve further investigation. For papers with cross-disciplinary impact in fact, the value of a citation is negatively related to the length of the reference lists. A citation by a publication in physics would value less than that by a paper in mathematics simply because the former's list of references is in general longer.

The great majority of the more popular bibliometric indicators and the rankings based on their use present two fundamental limits: lack of normalization of the output value to the input value, and absence of classification of scientists by field of research. Without normalization there cannot be any measure of productivity, which is the quintessential indicator of performance in any production unit; without providing field classification of scientists, the rankings of multi-field research units will inevitably be distorted, due to the different intensity of publication across fields. An immediate corollary is that it is impossible to correctly compare productivity at international levels. To date in fact there is no international standard for classification of scientists and, we are further unaware of any nations that classify their scientists by field at domestic level, apart from Italy. The commonly accepted definition of productivity, i.e. the number of publications per researcher, makes little sense, because publications have different values. We have proposed a proxy measure of productivity, FSS, which embeds both quantity and quality of production, and permits measurement at different organizational levels. Both the indicator and the related methods can certainly be improved, however they do make sense according to economic theory of production. Other indicators and related rankings, such as the simple number (or fractional counting) of publications per researcher, or the average normalized impact, cannot alone provide evaluation of performance - however they could assume meaning if associated with a true measure of productivity. In fact if a research unit achieves average levels of productivity this could result from both average number of publications and meaningful impact, but also from the opposite case of high numbers of publications and low impact. In this case, knowing the performance in terms of number of publications and average normalized impact would provide useful information on which aspect (quantity or impact) of scientific production to strengthen for betterment of production efficiency.

Aside from having an indicator of research unit productivity, the decision-maker could also find others useful, such as ones informing on unproductive researchers, on top researchers (10%, 5%, 1%, etc.), highly-cited publications, dispersion of performance within and between research units, etc.

For the large part of the objectives and contexts where evaluation of research performance is conducted, productivity is either the most important or the only indicator that should inform policy, strategy and operational decisions. We thus issue a two-fold call to the scholars in the subject: first, to focus their knowledge and skills on further refining the measurement of the FSS indicator in contexts of real use; second, to refrain from distribution of institutions' performance ranking lists based on invalid indicators, which could have negative consequences when used by policy-makers and research administrators.